\documentclass[conference]{IEEEtran}
\IEEEoverridecommandlockouts
\usepackage{cite}
\usepackage{amsmath,amssymb,amsfonts}
\usepackage{algorithmic}
\usepackage{graphicx}
\usepackage{textcomp}
\usepackage{hyperref}       
\usepackage{xcolor}

\def\BibTeX{{\rm B\kern-.05em{\sc i\kern-.025em b}\kern-.08em
    T\kern-.1667em\lower.7ex\hbox{E}\kern-.125emX}}

\begin{document}

\title{QoS-based Trust Evaluation for Data Services as a Black Box\\

\thanks{This work has been done in the context of the project SUMMIT (\url{http://summit.imag.fr}) funded by the Auvergne Rhône Alpes region.}
}

\author{\IEEEauthorblockN{Senda Romdhani}
\IEEEauthorblockA{\textit{Univ. of Lyon, CNRS} \\
\textit{Univ. of Lyon 3, LIRIS} \\
Lyon, France \\
senda.romdhani@univ-lyon3.fr}
\and
\IEEEauthorblockN{
Genoveva Vargas-Solar}
\IEEEauthorblockA{\textit{French Council of Scientific Research (CNRS)} \\
\textit{LIRIS} \\
Lyon, France \\
genoveva.vargas-solar@liris.cnrs.fr}
\and 
\IEEEauthorblockN{
Nadia Bennani}
\IEEEauthorblockA{\textit{Univ. of Lyon, CNRS} \\
\textit{INSA-Lyon, LIRIS} \\
Villeurbanne, France \\
nadia.bennani@insa-lyon.fr}
\and
\IEEEauthorblockN{
Chirine Ghedira-Guegan}
\IEEEauthorblockA{\textit{Univ. of Lyon, CNRS} \\
\textit{iaelyon - Univ. of Lyon 3, LIRIS}\\
Lyon, France \\
chirine.ghedira-guegan@univ-lyon3.fr}
}

\maketitle

\begin{abstract}
This paper proposes a QoS-based trust evaluation model for black box data services. 
Under the black-box model, data services neither export (meta)-data about conditions in which they are deployed and collect and process data nor the quality of data they deliver. Therefore, the black-box model creates blind spots about the extent to which data providers can be trusted to be used to build target applications. The trust evaluation model for black box data services introduced in this paper originally combines QoS indicators, like service performance and data quality, to determine services trustworthiness. The paper also introduces \textbf{DETECT}: a Data sErvice as a black box Trust Evaluation arChitecTure, that validates our model. The trust model and its associated monitoring strategies have been assessed in experiments with representative case studies. The results demonstrate the feasibility and effectiveness of our solution.

\end{abstract}

\section{Introduction} \label{sec:intro}
The explosion of data collected, managed, and provided using online services leads to the logic of "Everything as a Service" (XaaS: X as a Service).
Currently, the use of data as a service for accessing large volumes of data from different providers concerns mainly data consumers who see services as practical and easy-to-use components for accessing data.  

In this context, searching data implies looking up for data services that can potentially provide them according to specific conditions and quality requirements (freshness, availability etc.). However, reliance on services for data access requires them to be trustworthy and of good quality. This need is even more crucial when we are in the case of data integration\cite{a3} as data is being provided and composed using multiple data services presenting different quality of services and distinct data quality. 
By trust, we mean the belief that a service meets the terms and quality of service as promised by the service provider before use. 

In the case of data as a service (i.e. data services), this trust is related not only to the technical behaviour of the service in terms of the way it deals with requests including its QoS (e.g., availability, response time, etc.) but also in terms of the conditions in which it provides data (freshness, update frequency, delivery delay). 

Service level agreements (SLAs) contain clauses about the promised QoS for a given service along with penalties that detail acts to be made in case the service provider violates these QoS clauses. Service providers often deploy solution to monitor the functioning of their services (i.e. service performance) by computing QoS measures\cite{p1}. The monitoring is performed to make internal technical decisions related to resource allocation and so on. However, for privacy and security reasons, service providers rarely share their service monitoring measures. Thus, they do not provide guarantees to service users that service performance adhere to the promised QoS in SLA.
Furthermore, SLA does not yet contain clauses about the promised data quality factor and thus, users do not know what to expect regarding this factor.
In this sense, data services adopt a black-box model to export their API (Application Programming Interface) with methods and Input/Output parameters specifying their data types. 

Under the black-box model, services do not provide details on their backend, including how data are collected, updated and to which extent data are fresh (i.e. up-to-date). Still, clients are looking for alternatives to evaluate the trust level of both services and data to use it to choose services for building information systems and retrieving data.

To this end, we deem necessary providing a trust measure for black box data services to applications using them. This trust measure is computed using service performance and data quality factors. Therefore, the challenge is to propose (1) a trust model that captures and combines service performance and data quality aspects in order to meet tightly applications needs and (2) a mechanism to collect information required to compute services  trust level, in case there is no or little access to performance and data quality meta-data.

In order to illustrate the aspects to consider behind this challenge, let us consider an e-health scenario
\footnote{This scenario was proposed in the context of the project xx to be precised later} 
that highlights the importance of considering trust when looking up for data services that can provide the right data for a target application.

Fever in chemotherapy-induced neutropenia (FN) is the most frequent, potentially lethal complication of chemotherapy in pediatric and adult patients with cancer \cite{a1,a2}. 

Fever is particularly concerning if it occurs when the white blood count level is known to be low. During this time, the body’s normal defences against infections are expected to be down. 
To this end, fever is considered as a medical emergency since it indicates a possible infection during chemotherapy. Therefore,  cancer patients' body temperature should be monitored continuously.

Alice is a 50-year-old woman with breast cancer treated with chemotherapy. To avoid infection complications when she is at home, she uses medical devices provided by the hospital's emergency service to keep track of her temperature and other physiological measures. 

Devices are programmed to capture data according to some unknown rate. According to a specific update frequency (i.e. insertion frequency),  measures are sent to the hospital that maintains a database with Alice's sensed data. To complete this sensing procedure, Alice uses a connected thermometer to measure her temperature frequently.  The device sends the information directly to an e-health service installed in her smartphone. She also records her temperature using a regular thermometer every x days or weeks. Alice keeps track of her manually measured temperature in a specific service installed in her smartphone too.  The manual temperature measurements frequency is variable as it is determined by  Alice's mood, memory to measure the temperature, whether she is alone and whether someone else comes by and encourages her to do a check-up. 

Alice gives access to all her collected data through the services that are deployed on her smartphone. Doctors should be able to access her data at any moment. This data must be reliable to be used effectively to make accurate decisions, and it must be delivered promptly in case of an emergency. 
Deciding which service to use to retrieve Alice's physiologic data in a given situation is challenging as services have different trust levels. Indeed, each service ensures different QoS and data quality guarantees since data are produced and updated under different conditions (production rates, update frequencies that affect data freshness).

In case of an emergency, choosing the right service to provide the best quality data can be critical for doctors. Therefore, in ideal conditions, services must be tagged with a trust measure (updated regularly) that could be used as data services' selection criterion. 

Our work addresses two research problems (RPs) for measuring data services trust level to support data lookup and selection where trust is critical.\\
(RP1) What is the appropriate model for describing data services QoS-based trust? How to define both service performance metrics and data quality metrics to be used in the trust evaluation process? \\
(RP2) How to collect the necessary information for this trust evaluation model?  


The main contribution of our work is a data service trust model combining performance (availability, task success ratio and response time) and data quality metrics (data and services' database timeliness).

Our proposal for estimating data services trust has been implemented  by DETECT a Data sErvice as a black box Trust Evaluation arChitecTure and validated using an e-health scenario. 
The scenario consists of data services giving access to HAPI-FHIR\footnote{HAPI FHIR is a complete implementation of the HL7 FHIR (Fast Healthcare Interoperability Resources) standard for healthcare interoperability and data exchange developed in Java, published by HL7\url{https://hapifhir.io}} servers on top of which trust metrics are observed and measured. DETECT's general principle is to continuously collect and monitor technical and data quality meta-data so that data services are tagged with an up-to-date trust measure. The experimental results show how the trust of services with different behaviour can be computed and used to rank them given services lookup requests.


The remainder of this paper is organized as follows. Section \ref{sec:related} enumerates related work addressing data and service trust solutions. Section \ref{sec:trustindex} describes our proposed QoS-based trust model for data services while defining the necessary performance and data quality metrics. Section \ref{sec:arch} presents the description of the used system DETECT for implementing our proposed trust evaluation model. The experiments to prove our solution's feasibility is presented in section \ref{sec:exp}.  The last section concludes the paper and discusses future work. 

\section{Related Work}\label{sec:related}
%

Service trust evaluation is the subject of many proposals in the literature that treat trust estimation issues in different service-based environments, including SOA, web services, cloud services, etc. Reading through the related work, we summarize the existing proposals focusing on trust: at (i) the service level, (ii) and the data level.

    \subsection{Trustworthy Services}
Existing service trust solutions differ in the type of the adopted trust assessment model (fuzzy models \cite{a7,a8,a9}, probabilistic models \cite{a10}, machine learning models \cite{a11,a12,a13}, multi-criteria decision-making models (MCDM) \cite{a14,a15} and classical mathematical models such as weighted-addition \cite{a16}).
    The service trust evaluation can be subjective \cite{a17,a18,ref_journal16}, objective\cite{a16,a19} or hybrid \cite{a20,p15} (subjective + objective). Subjective evaluation relies on user preferences and feedback about the service usage for evaluating trust. Users' feedback may be delivered as a quantitative opinion or as a qualitative opinion in textual form.
    Objective evaluation, on the other hand, involves measurable information such as the service capacities, service capabilities, and service performance \cite{autocite}.
    Note that users' subjective feedback may be influenced by several factors and may exhibit significant variations in the evaluation of the same service from one user to another. Besides, there is no guarantee of users' honesty. 

    As for performance, it can be measured using several metrics: service availability,  service response time,  service reliability, and service security.
    Thus, objective service trust evaluation mainly uses service performance \cite{autocite} as a reference measure which is also considered a significant criterion for selecting services. Accordingly, services performance is computed through SLA monitoring, where information is collected from previous experiences and is used to measure the degree of SLA fulfillement \cite{ref_proc4}. If the SLA is satisfied, then the service is considered trustworthy. 

    \subsection{Trustworthy Data}
        
     Data similarity and data provenance are the two most commonly used characteristics for assessing data trust. Similarity-based solutions consider that the  more similar the data  values from the same event are, the more likely the data is correct and therefore trustworthy. Provenance-based solutions considers that data with good provenance are more likely to be trustworthy.

     In addition, several studies focus on the definition of metrics for data quality evaluation\cite{p12,p16,p17,p18}. Authors assume that data providers already export this information as evidence of honesty or that it is accessible. 
    
    Authors in \cite{p6} evaluate data quality and data trustworthiness using the web provenance information.  Their approach can collect provenance information, evaluate it, and use it for data quality evaluation.
    
    
    Works proposed in \cite{p8,p9} evaluate location-based data trustworthiness using data provenance. Their approach employs GPS data to trace people’s location to generate meta-data about provenance available and accessible. Thus, they apply data similarity from different provenances to verify their trustworthiness.

    In \cite{p10,p11}, authors propose a systematic method for assessing the trustworthiness of data items in sensor networks. Accordingly, a trust score is computed using data provenance and their values based on their similarity "the more similar values for the same event, the higher the trust scores". In this approach, the trust score of the data affects the trust score of the network nodes that created and manipulated the data and vice-versa.
    
    The work in \cite{p20} introduces a novel data provenance trusted model in cloud computing. Authors believe that to establish trust with users, cloud service providers should consider providing trust information such as provenance record. Based on this hypothesis, meta-data about provenance is provided by a cloud provider.
    
    The work in \cite{p21} proposes trustworthy and secure management of provenance in cloud computing. It captures the complete provenance of any entities in the cloud without trusting cloud providers. In this proposal, the authors do not take into consideration the fact that provenance is meta-data hidden behind the service's API.
    
   
   In \cite{p18,p19}, authors propose a framework for data freshness evaluation in a data integration system (DIS). This framework assesses data freshness value changes in DIS for data moving from data sources to users. Besides, it evaluates the evolution of data freshness using graphs representing the integration workflows in a controlled environment. Thus, the proposed framework does not require meta-data about data provenance.

\subsection{Trustworthy Data Services: Discussion}
Existing work proposes models to define data quality metrics, including data freshness and performance metrics, service response time, task success ratio and availability. 

Certain solutions propose services trust evaluation models taking into consideration the performance factor. Those models are often not proposed for a specific type of service. In the cloud environment, solutions were proposed for services deployed at the "software a service" (SaaS) and "infrastructure as a service" (IaaS) layers. These trust solutions generally differ in the used trust factors and trust metrics and are not defined with the same formulas. To the best of our knowledge, no trust evaluation solutions targeting precisely data services combining QoS and data quality were proposed.  There is also no solution considering data freshness in their data trust evaluation models.
Nevertheless, they do define data quality metrics, including data timeliness.
Data quality metrics are evaluated in these works, assuming that the necessary information and meta-data are available for this evaluation. The general assumption is that data providers may export data quality observations as proof of goodwill to cooperate and show that they provide "good" data.
In the absence of information, those measures need to be defined differently to evaluate data freshness, especially database timeliness.

\section{Data Service Trust Evaluation Model}\label{sec:trustindex}

Service providers propose services under heterogenous quality of service (QoS) conditions to support various users needs related to scalability, fast response time, availability etc. QoS varies from one service provider to another, from the same service provider at different moments (e.g., from day to day), or even from different applications according to the application domain. As a result,  services have different associated QoS levels. However,
there are no guarantees that the  QoS is continuously ensured, and thus,  it is necessary to provide measures that can serve to determine to which extent a certain level of QoS can be guaranteed and for how long. Besides, when services are data providers, it is also necessary to determine data reliability.
%
%
As mentioned in the e-health scenario, data reliability is determined by different factors, including data freshness.

A trust measure can 
be a representative indicator about services QoS and data quality used as a quality reference by services consumers. 
Therefore, a high trustworthiness level of a given service can be a representative indicator of the performance and data quality standards followed by a service and satisfies the quality requirements of service consumers.

To this end, when we seek a data service from a provider, it is relevant to consider two factors. First, we consider past experiences with the data service, resulting in an overview of its performance level \cite{a13}. Second, we look at the quality of the data delivered by the data service. Thus, the trust level  of a data service is based on two factors: \textit{performance} 
and \textit{data quality}, and it is defined in [0,1] as follows:
\begin{equation}
    T_{DS}=\alpha* Performance+ \beta *Data Quality
    \label{teq}
\end{equation}

Where $\alpha, \beta$ weights pondering the performance and data quality factors are described in the following subsection. These ponder values vary according to the importance given to each of these factors by an application.

\subsection{Performance Factor}
\textit{Performance} describes the computing capacity of a data service
defined by three metrics: availability, time efficiency, and task success ratio. The general idea behind the performance factor is that the data service is expected to be available when it is requested, it should adhere to the response time described in its SLA, and it must systematically deliver data to consumers successfully. 
\begin{equation}
    Performance = \sum W_j * Q_j   
    \label{perf}
\end{equation}
Where \textit{$Q_j$} = \{availability, time efficiency, task success ratio\}. \textit{$W_j$}: weight of the metric j, which varies according to the importance of a metric in a given context to the data service's consumer. 

\begin{itemize}
    \item \textbf{Availability (Av):} 
    A data service is unavailable when a  request is denied  \cite{a16,p14}. For instance, doctors in the e-health scenario must be sure that data services are available whenever they need to access Alice's data.
    
    Availability can be defined as the degree to which a data service is operational and accessible when requested. Therefore, we define availability in [0,1] as follows:
    \begin{equation}
    Av=\frac{A_k}{N_k}  
    \end{equation}
    Where $A_k$ is the number of accepted requests by the data service k and $N_k$ is the total number of requests submitted to the data service k. 
    
    \item \textbf{Task Success Ratio (TSR)}:
    Sometimes, data services may be available and accessible but are unable to deliver data to data consumers due to
    network failures, time outs set by the consumer, etc.
    The task success ratio of a service measures successful data delivery in response to accepted requests. Back to our scenario,  a low task success ratio associated with a given data service indicates to doctors that there is a considerable risk of not getting the needed information.

    \begin{equation}
    TSR=\frac{S_k}{A_k} 
    \end{equation}
    Where $S_k$ is the number of successful requests where data arrived to destination by the data service \emph{k}. TSR is in [0,1]. 
   
    \item \textbf{Time Efficiency (TE):} 
    quantifies how well a data service meets the expected response time (ERt)  promised by the service provider. 
    Indeed, having a data service with a low response time violation on average w.r.t ERt indicates that one can rely on this service for urgent decision making more than a service with a higher response time violation. 
    Time efficiency metric is defined in [0,1] as follows:
    
    \begin{align} 
     TE =&\quad 1 - \frac{Rt}{ERt} &\qquad \textrm{if} \qquad  Rt < ERt\\ 
     TE =&\quad 0  &\qquad \textrm{if}\qquad  Rt > ERt
    \label{DT}
    \end{align}

    where $Rt$  is the response time of the service on average. 
   
\end{itemize}

\subsection{Data Quality Factor}
Knowing the quality of 
data delivered by data services
helps to determine the reliability of data and its fitness-for-use. 
According to the literature \cite{autocite}, data quality can be evaluated using multiple data quality dimensions including completeness, timeliness, accessibility, interpretability etc.

In our work, we estimate data \textit{data freshness} because we apply our results to applications for which data must be up to date, for example e-health applications.

\textit{Data freshness} indicates the extent to which data is fresh in the sense that it is meaningful for a target application \cite{p5,p19}. The principle behind this is that fresh data is valuable and trustworthy compared to outdated data that loses value and can negatively affect decisions made using it.


Data freshness is measured using the notion of  \textit{timeliness} 
covered by two dimensions: \emph{data timeliness} measuring the extent to which data is up-to-date; and \emph{database timeliness} measuring how often data is inserted into a database.
Intuitively, the idea is that "data is made available as quickly as necessary to preserve its value" \cite{p2,p7}. 
%
Thus, data is said fresh when it is up-to-date and inserted frequently into a database. Let $T_D$ and $T_{DB}$ denote respectively data timeliness and database timeliness for a given data set. 

We assume that data are produced under different production rates $P_R$ that indicate how often data is collected (i.e. produced), for example, from the "real world" and timestamped when produced. Data production rate may be static (i.e., it does not change over time) or programmed to be triggered upon the production of specific events. 
For instance, in the e-health scenario, the smartphone or smartwatch can be programmed to collect temperature when Alice is active or far from home.

We suppose that data remains fresh within a specific time interval $T$ called the validity interval. $T$    is predefined according to the application domain. 

Assuming that data producers continuously send data to  a data service with a specific update frequency $U_f$. 
\begin{itemize}
        \item \textbf{\textit{Data Timeliness}} captures the gap between the production of data and the time when it is needed and makes sure it is produced within the data validity interval $T=$[$t_{min},t_{max}$]. 
     
    We define timeliness for data D in [0,1] as follows:
    \begin{align} 
     T_D =&\quad 1-\frac{t_R-t_P}{T} &\qquad \textrm{if}\qquad  t_R < t_{max} \\ 
     T_D =&\quad 0  &\qquad \textrm{if}\qquad  t_R > t_{max}
    \label{DT}
    \end{align}

    Where $t_R$ represents the request time, $t_P$ is the data production time, $t_{max}$ is the maximum time for data to be fresh and $t_{min}$ = $t_P$.
    The closer $T_D$ is to $t_{max}$, the less the data is considered timely, that is, the less fresh it is.  Beyond $t_{max}$, data in no longer fresh.
    
    For example, using our scenario, data timeliness indicates to the doctors that the data service managing home thermometer measures might provide outdated or stale temperature readings that might not be accurate concerning the actual patient's health state. 
    \begin{figure}[!htbp]
    \includegraphics[width= \linewidth]{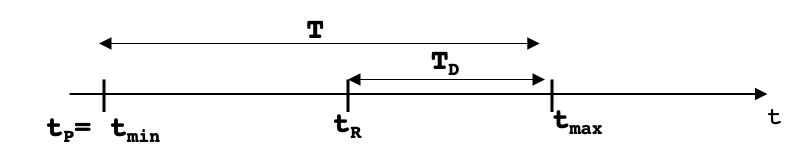}
    \caption{Data Timeliness.}
    \label{archi}
    \end{figure}
   
    \item \textbf{\textit{Database timeliness}} measures how often a database is updated given by the database update frequency. 
     We say that a database is updated when new data are inserted. The intuition is that frequent updates can contribute to the preservation of data freshness. 
     The frequency of insertions gives the database timelines. For black-box services, this frequency must be "guessed" by comparing data obtained by successive requests. Thus, the database timeliness is not stated by a formula but through a protocol based on statistical and heuristic analytics strategies. The protocol concerns our ongoing work, and it is out of the scope of this paper.
    

\end{itemize}

From the above definitions of data quality metrics, it is possible to have an intuition concerning a  possible correlation among the data freshness metrics. 
For example, if a database is updated frequently, it is more likely to be timely, but not necessarily the contrary. If the database has not been updated within a data validity interval (stated by the application domain of data consumers), data is more likely to be outdated from the data consumer perspective. Thus, we define data quality in [0,1] as follows:


\begin{equation}
    DataQuality=T_D * T_{DB}
    \label{edq}
\end{equation}

        

\section{DETECT: General Description }\label{sec:arch}

Figure \ref{archi} shows the general structure of DETECT that implements our trust evaluation model for black-box data services.
The main objective of DETECT is enabling data requesters to select the most reliable data service according to their needs providing a trust-sorted list of data services tagged with their up-to-date trust index. 
DETECT is composed of three main modules: (i) the performance measuring module (PMM); (ii) the data quality measuring module (DQMM); and (iii) the trust measuring module (TMM). 
PMM monitors data service's performance, DQMM  observes and collects relevant metrics for evaluating the data quality provided by a data service, and TMM collects both data quality and performance measurements and computes data services trust scores. 

\begin{figure}[!htbp]
\includegraphics[width=1 \linewidth]{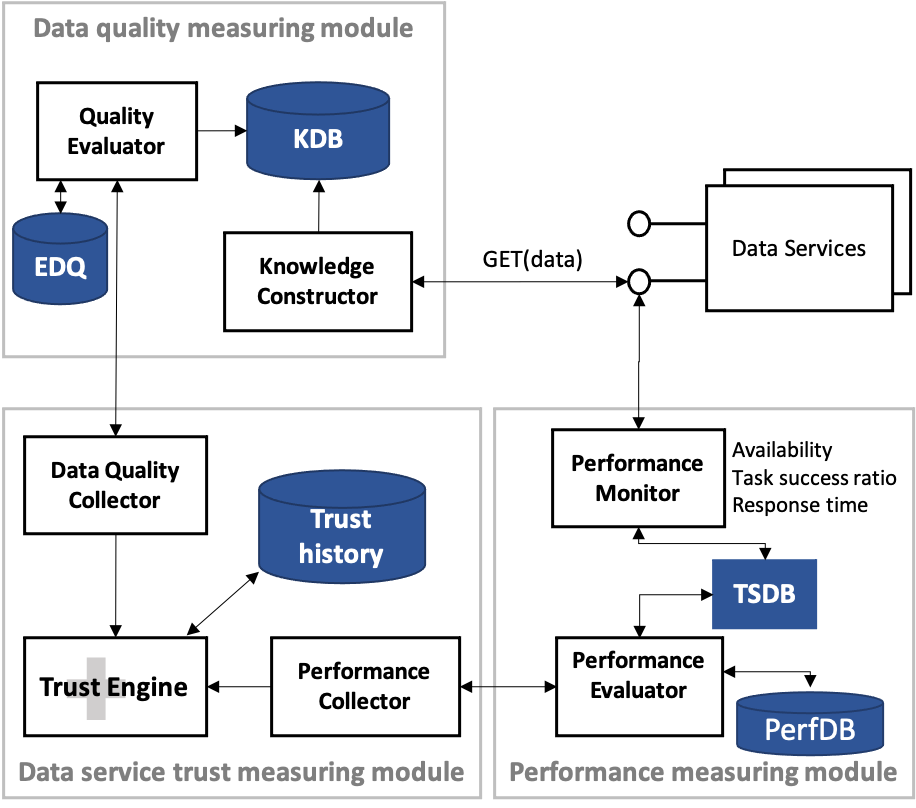}
\caption{Data Service Trust Evaluation Architecture.}
\label{archi}
\end{figure}

\subsection{Performance Measurement Module}
The PMM performs its operations in two phases: monitoring and evaluation:
\begin{itemize}
    \item During the monitoring phase, the \textit{Performance Monitor} scrapes continuously the performance metrics using the service's API including data service response time, task success ratio and availability and record for a given instant (i.e. timestamp) their values. 
    
    Several interesting tools for scraping and observing such performance metrics have been proposed in the literature \cite{ref_journal15} and are available on the internet (i.e. JMETER, Prometheus etc.).

    The \textit{Performance Monitor} stores performance measurements in a time-series database (TSDB) which indicates the recorded response time of services in milliseconds, whether the service was available and whether it successfully finished the job for each scraping timestamp. 
    
    \item During the evaluation phase, the \textit{Performance Evaluator} creates a performance level base by collecting evidence from the TSDB. 
    
    First, it collects performance measurements made within a specific time interval including response times for the different timestamps, the number of requests made, the number of accepted requests and the number of successful requests within this time interval. 
    
    Second, it computes performance metrics as defined in \ref{sec:trustindex} by computing (1) the reponse time on average for the corresponding time interval, (2) the availability and (3) the task success ratio for each service. 
    
    Third, the \textit{Performance Evaluator} evaluates the performance level of the corresponding service by applying equation \ref{perf}. The \textit{PerfDB} database stores the list of services tagged with their evaluated performance level along with the evaluation timestamp. This list is updated periodically each time the performance of the service is evaluated.  
\end{itemize}
an API is made available to enable the trust measuring module to access the latest recorded performance for the available services. 
\subsection{Data Quality Measurement Module}

DQMM observes and evaluates the quality of the data accessed by the corresponding data service. This evaluation is based on both data timeliness and database timeliness defined in section \ref{sec:trustindex}.

The assessment of\textit{ data timeliness} and \textit{database timeliness} can be achieved from the knowledge about data sources and meta-data about data quality, including data production time and database update frequency. As aforementioned, we suppose that data is timestamped. However, the database update frequency is unknown to the data service's user, and the black box character of these data services creates abstraction about how data are captured and processed.  

A resource-consuming data quality evaluation process must be set up to provide timeliness measures.
Thus, we propose defining and deploying a meta-data observability protocol addressing issues related to the absence of the necessary meta-data for data quality evaluation, especially the database timeliness evaluation. 

The protocol follows a series of steps.
First,  the \textit{knowledge constructor} observes the database state changes of a given data service as a black box (DSaaB) using sampling techniques. Changes in which we are interested are the INSERTs statement performed of the service's related database. 

By pulling data samples continuously with a specific sampling frequency, whether it is random or static, the \textit{knowledge constructor} observes and measures some metrics that help build knowledge about the data service's change history. The sampling frequency must be chosen in such a way as to find a trade-off between retrieving enough representative samples for the measurement of those metrics and reducing the overhead produced by sending requests and processing data.

For \textit{data sample timeliness}, we apply the equation \ref{DT} on every data item in the sample and then compute the average for all the data set. For the \textit{database change of state}, one would compare the different database states such as that we have a $\delta \neq \{\}$. $\delta$ represents the intersection between two data sets belonging to two consecutive samplings. Thus, the approximate update frequency of the database managed by a data service can be measured.
The database \textit{KDB} stores those metrics along with the observation's timestamp.  

Using \textit{KDB} and per observation period, the \textit{quality evaluator} (1) measures database update frequency, (2) computes data timeliness on average and (3) uses those last two metrics to evaluate the data timeliness and thus the data freshness of the data source in question. 
The result of this module is a list of data services tagged with their measured data quality level stored in the database \textit{EDQ}. Data quality is measured periodically in order to keep this list up-to-date.

An API is made available to enable the trust measurement module to access the latest recorded data quality for the available services.
\subsection{Data Service Trust Measuring Module}

The Trust Measuring Module  (TMM)  evaluates a data service's trust level using the PMM and DQMM results. Thus, two collectors are used in DETECT: the \textit{performance collector} which collects performance levels of the available data services; and the \textit{data quality collector} which collects data quality levels of the available data services. These collectors gather trust factors and feed the \textit{trust engine} on-demand. The trust engine evaluates the trust level of each available data service using equation \ref{teq}. It further enables the data requester to specify the degree of importance of trust factors, including performance and data quality. 
A history of the evaluated trust levels of each service for the different requests is stored in the  \textit{trust history} which contains for every measuring timestamp the service ID, its data quality level, its performance level and the corresponding data service trust level. 

\section{Implementation \label{sec:exp} }

We implemented a prototype that realizes the QoS-based trustworthy data services selection. In the following, we present the implementation details using our e-health scenario; we describe two case studies and the experimental setting to show the feasibility of our solution. 

We used DETECT to set up Alice's homecare scenario for temperature monitoring. We used Python to develop the different modules of our architecture (i.e., PMM, DQMM and TMM). We use SQLite to implement the database integrated in the architecture. \footnote{\url{http://sqlite.org}}.

We used Prometheus\footnote{\url{https://prometheus.io}}, a monitoring tool, and its related JMETER exporters\footnote{\url{https://jmeter.apache.org}} to implement the \textit{Performance Monitor} and to continuously observe  availability, response time and task success ratio metrics. 


We implemented data services using HAPI FHIR solutions which are built from a set of components called \textit{Resources} used to exchange and$/$or store data and can be easily assembled to solve a wide range of healthcare-related problems. A resource is an entity that (1) has a known URL by which it can be accessed, (2) contains a set of structured data items and (3) has an identified version that changes if its contents change.

We mainly measured the quality of data accessed through the resource \emph{Observation}\footnote{\url{https://www.hl7.org/fhir/observation.html}} that is a central element in healthcare, used to support diagnosis and monitoring progress. 
%
Uses of the Observation resource include (1) \textit{vital signs} such as blood pressure and temperature, (2) laboratory data like blood cells count, (3) device measurements such as EKG data or Pulse Oximetry data, and so on. Therefore, we can develop our e-health scenario using this resource for monitoring temperature.
HAPI FHIR resources are queried using RESTful APIs (GET, POST, PUT,  DELETE, UPDATE). 

The scenario runs on a 64GB macintosh Macbook Pro where data services are deployed on  self-contained environments using  DOCKER\footnote{\url{https://www.docker.com}}. Containers were configured with different performance levels  managed through the number of CPU cores and the memory allocated to each container.  

\subsection{Trust case studies}

This section presents two case studies based on the eHealth scenario to validate the trust model and DETECT's applicability. Each case study is designed to target one or both trust factors, including performance and data quality. We assume that we have two kind of services: on-demand services and pushed services.
The chosen case studies illustrate various applications requirements for which performance and data quality will be weighted accordingly. 

\begin{itemize}
    \item 
    \textbf{Case Study 1: Alerting} 
    
    This case study addresses continuous monitoring of Alice's temperature. In the case of a significant variation in Alice's temperature, the devices' push service must immediately alert Alice's doctor. For instance, sending a message with her current temperature and location.  Both performance and data freshness are essential since up-to-date data must be transmitted immediately when a sudden variation in temperature occurs. Since Alice's safety is at risk, data must be sent timely, and thus the performance criterion for selecting services seems more critical than data freshness. 
    Therefore, more weight is attributed to the performance factor when evaluating the trust levels of the available and pre-selected services using the equation \ref{teq}. 
    
    \item 
    \textbf{Case Study 2: Instant checkup by Alice's doctor} 
    
    In this case study, we assume that Alice's doctor wants to perform from time to time an instant checkup on her health latest indicators using on-demand services. For example, she wants to have the latest recorded temperature, and thus data must be fresh. As described in our scenario,  temperature readings are done manually and digitally using different devices (e.g., hospital's devices, her connected thermometer and her manual thermometer), and measures are taken at different frequencies and rates.
    This use case emphasises the data quality factor (i.e., data freshness produced by services managing the data produced by the different devices).
    Thus, one  expects the system gives higher weight to data quality factor, when evaluating the services trust using the equation \ref{teq}.

\end{itemize}
\subsection{Experimental Setting}
In the case studies described above, three HAPI FHIR servers with different FHIR standards have been deployed on docker containers: one simulating the hospital's server, one simulating Alice's smartphone' server and the last one simulating the SOS server of the hospital. Each server has its independent database on the corresponding server and is reachable through its own URL.  

In order to give access to these servers, we deployed 7 data services each with its own access point: three data services giving access to the three devices used by Alice are deployed on the first two HAPI FHIR servers. Only one service giving access to the third HAPI FHIR server. 

\begin{figure}[!htbp]
\centering

\includegraphics[width=.8 \linewidth]{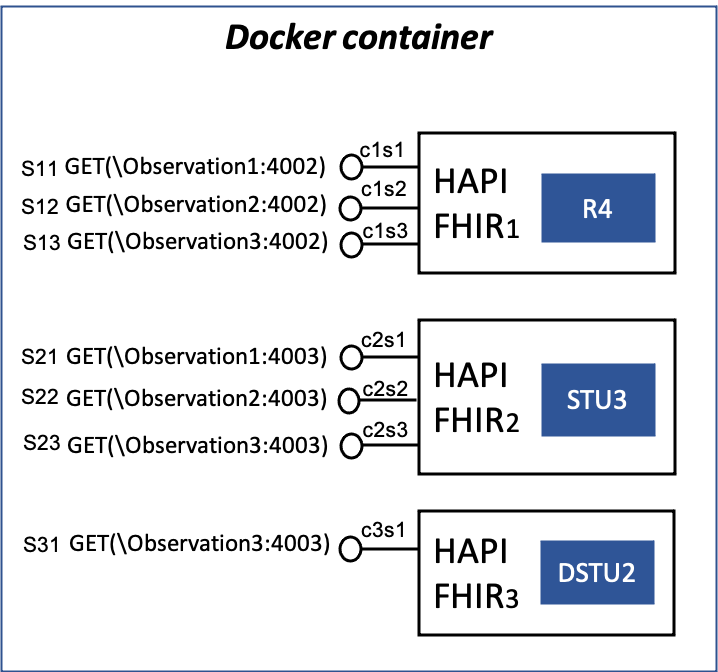}
\caption{Experimental setting: Docker containers.}
\label{dc}
\end{figure}

Note that these servers are configured the same way in our controlled environment.
However, to simulate the variation in the performance of the different data services, we allocated different number of CPU cores and different number of services to the three HAPI FHIR servers in a way that: (1) The bigger the number of the allocated CPU resources, the better the performance of the service since we have more resources for the server 
and (2) The smaller the number of the deployed services on the same container, the better the performance since the available resources on the server are shared between a smaller number of services. 

1 CPU core is allocated to Alice's smartphone' server, 2 CPU cores are allocated to the hospital's server, and 2 CPU cores are allocated to the SOS server (see figure\ref{dc}). By allocating CPU, we limit the performance of the different services. Following this logic, service S31 has the highest performance since it is deployed solely on the container with 2 CPU cores followed by the three services (S21, S22, S23), which are deployed on the second container and also have 2 CPU cores and finally the three services (S11, S12, S13) on the first container have the worst performance rate since they have only 1 CPU core compared to the second container. 

The performance metrics are observed and measured using a configurable thread group with the help of JMETER exporters. A thread group is a set of processes executing the same job. JMETER enables us to set the number of processes of a thread group to access a given service's URI simultaneously. Each thread simulates a user request to a data service. JMETER also enables us to set the duration for all users in the thread group to execute all processes. 
For instance, if a thread group has 20 processes and the period to access the service is set to 40 sec, JMETER will take precisely 40 sec to start all processes (i.e. users), and thus a process is started every 2 sec. The observed performance metrics are scraped by Prometheus every 15 sec and stored in its TSDB. The \textit{Performance Monitor} has access to this TSDB every 300 sec to compute the performance of the related data services.


In the absence of real-time real-world data services, we have simulated the process of data production and data insertion for Alice's three devices. To reduce the experimental duration and for the sake of simplicity, we have configured the data validity interval to 60 sec, which means that data about temperature remains valid for only 60 sec, and beyond this time, data is no longer considered fresh.

Also, we have varied data production rates and data insertion rates for all devices in a way that they have different data quality levels w.r.t our experimental setting: (1) all data services have the same data production rate (5 sec), (2) Data services S11 and S21 have the highest insertion rate equal to 20 sec (2 to 3 insertions per data validity interval), (3) followed by data services S12 and S22 with an insertion rate of 40 sec (1 insertion per data validity interval), (4) data services S13 and S23 with an insertion rate of 100 sec (1/2 insertion per validity interval) and (5) last data service S31 has the lowest insertion rate equal to 150 sec (1/3 insertion per validity interval). According to this configuration, it is expected that services S11 and S21 have higher data quality as they have the highest database timeliness, followed by S12 and S22, followed by S13 and S23 and finally S31.

Data insertion is performed using APIs of the deployed data services, and data about Alice's body temperature are made accessible through the resource observation.  
Table \ref{tab1} ranks services according to the expected performance and data quality levels.

\begin{table}[htbp]
\caption{Services Ranked According to Trust Factors}
\begin{center}
\begin{tabular}{|c|c|}
\hline
\textbf{\textit{Performance}}& \textbf{\textit{Data Quality}}  \\
\hline
S31  & S11, S21  \\
\hline
S21, S22, S23  & S12, S22 \\
\hline
S11, S12, S13  & S13, S23 \\
\hline
  & S31 \\
\hline
\hline
\end{tabular}
\label{tab1}
\end{center}
\end{table}

\begin{figure}[!htbp]
\includegraphics[width=1 \linewidth]{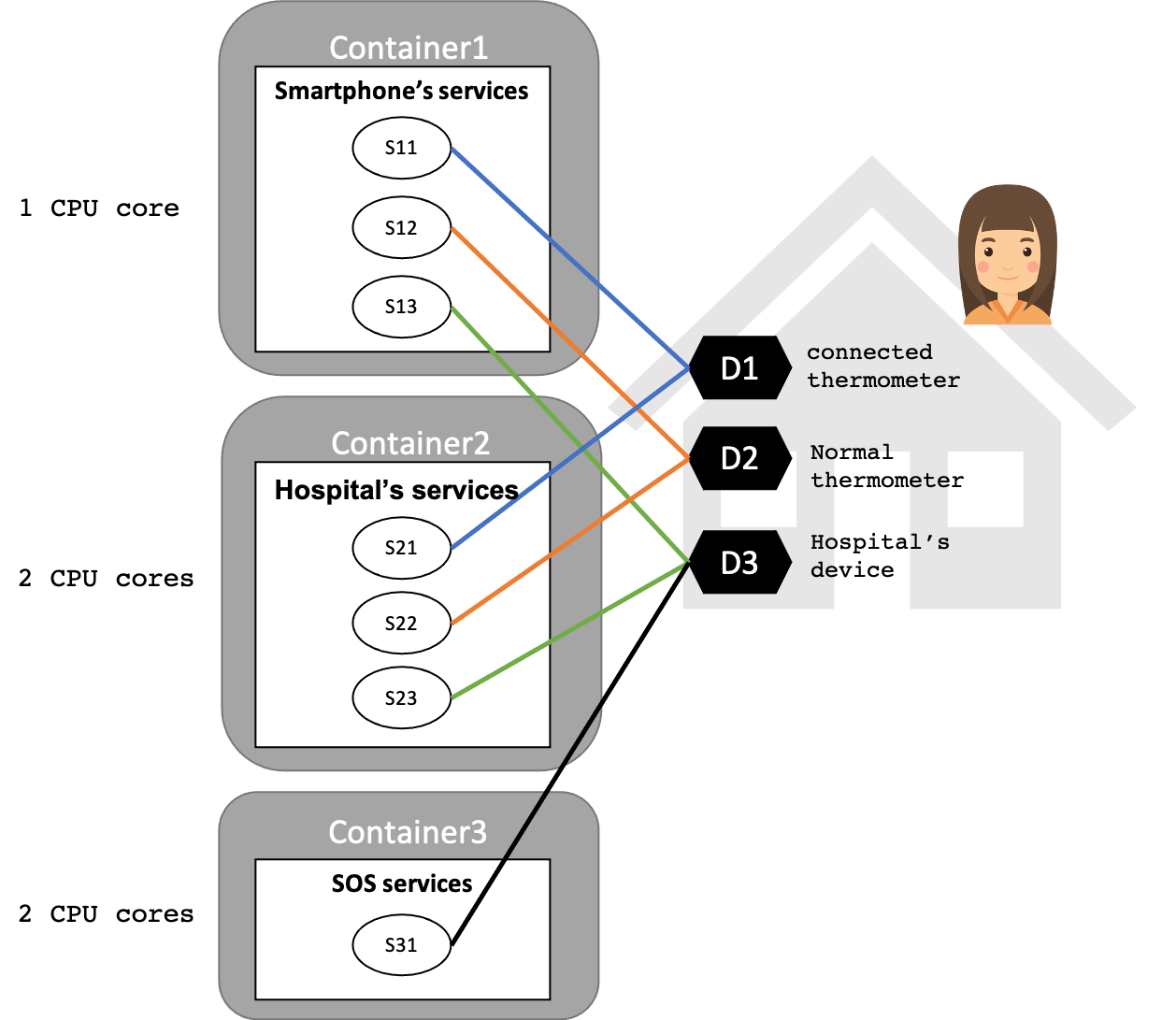}
\caption{Experimental setting: Data Service Trust Evaluation.}
\label{es}
\end{figure}

Once all the experimental setting is ready and to test our solution, we performed some trust requests to the \textit{Trust Engine}. As a result, the \textit{trust engine} outputs a ranked list of the available services according to their trustworthiness level. 

Our experiment consists of testing the effect of $\alpha$ and $\beta$ over the trust level of data services and their ranking: For each request, at a given time \emph{t} and for a fixed performance and data quality levels, we varied $\alpha$ and $\beta$ in order to show the effect of the evaluated data quality level and performance level on the trust level evaluation of data services. To do so, we decrease the weight $\alpha$ from 1 to 0 while increasing $\beta$ from 0 to 1 (see table \ref{tab2}).
    
    
Concerning performance evaluation at the \textit{Performance Evaluator} level, we currently consider availability, task success ratio and response time equally important, and thus they have equal weights when computing performance level as in equation \ref{perf}.



\subsection{Results and Discussion}

We performed tests using the configuration described above in order to verify our QoS-based trust model and architecture. 
As described for the test, we sent a trust request and observed the results. The resulting ranked lists of services are presented in table \ref{tab2}. 
\begin{table}[htbp]
\caption{trust Request Output}
\begin{center}
\begin{tabular}{|c|c|c|c|c|c|}
\hline
\textbf{\textit{$\alpha$=1,$\beta$=0}}& \textbf{\textit{$\alpha$=0,7,$\beta$=0,3}} & \textbf{\textit{$\alpha$=0,5,$\beta$=0,5}} & \textbf{\textit{$\alpha$=0,3,$\beta$=0,7}} & \textbf{\textit{$\alpha$=0,$\beta$=1}} \\
\hline
S31  & S22 & S21 & S21 & S21 \\
\hline
S23  & S21 &  S22 & S22 & S11\\
\hline
S22 & S31 &  S31 & S31 & S22\\
\hline
S21  & S23 &  S23 & S23 & S12\\
\hline
S13  & S11 &  S11 & S11& S31\\
\hline
S12  & S22 &  S12 & S12 & S13\\
\hline
S11  & S13 &  S13 & S13 & S23\\
\hline
\hline
\end{tabular}
\label{tab2}
\end{center}
\end{table}

According to table \ref{tab2}, we notice that:
\begin{itemize}
    \item The service S31 has the highest trust level when we only give importance to service performance ($\alpha$=1 and $\beta$=0); the less we give importance to performance, the more S31 is ranked lower in the list (5th place when $\alpha$=0). 
    \item The hospital's services (S21, S22, S23) are ranked below S31 when $\alpha$=1 and $\beta$=0. The more we give importance to the data quality factor, the better they are ranked and overpass the trust level of S31 excluding S23. 
    
    \item Services deployed on the first container (S11, S12, S13) are ranked last in the list when $\alpha$=1 and $\beta$=0. The more we give importance to the data quality factor, the better they are ranked and overpass the trust level of S31 excluding S13. No matter the weight we attribute to trust factors, services deployed on the smartphone HAPI FHIR server are consistently ranked below the services deployed on the hospital's HAPI FHIR server.
\end{itemize}

Hereafter, we discuss the results for each case study.

\begin{itemize}
    \item \textbf{Case study 1}: as aforementioned, in this  case study, we are only interested in selecting data services with the highest performance level: $\alpha$=1 and $\beta$=0. 
    
    According to the above observations and as shown in table \ref{tab1},  experimentations demonstrate the feasibility of our trust model and architecture since results provide the suitable trust-based ranking of data services as expected. Note that in our configurations, we control the performance of HAPI FHIR servers but not the performance of services deployed on the same server. These services run independently, and the allocation of resources among them depends on the docker containers' load balancing and scheduling method. However, for services deployed on the same container, we can perform some actions to diminish the performance of some of them, like sending more user requests.

    \item \textbf{Case study 2}: in this case study, we are  interested in selecting data services with the highest data quality: $\alpha$=0 and $\beta$=1. 
    
    According to the above observations: 
    (1) Services S21 and S11 have the best data quality. This result is not surprising because the configured data insertion rate for these two services is the highest. However, S21 is better than S11.
    (2) As expected, S12 and S22 are ranked next since they have a lower insertion rate. Still, S22 is better than S21. Services S13 and S23 followed them. (3) S31 was ranked before S13 and S23.
    
    These results can be explained by the correlation that exists between data quality and services' performance. The value of the data timeliness highly depends on the performance level of the corresponding service. The characteristics of the infrastructure negatively influence data timeliness. An infrastructure configuration providing limited resources punishes the service's response time, and in consequence, the data timeliness suffers too.  Note that this correlation affects the data timeliness value because data in our scenario changes frequently and has a very small validity interval (in matter of seconds). Thus, data looses rapidly its freshness: the longer the service takes to respond to the query, the worst the data freshness.
    For instance, services on the hospital's server are always better ranked than those deployed on the smartphones server even though their data production and insertion rates configuration is similar.
    
    The "bad" metrics of the smartphones servers are because the performance configuration of the hospital's server gives access to more resources.
    
    This dependence should be represented through the data service trust evaluation model defined in equation \ref{teq}. Nevertheless, we did not consider this dependence yet between the trust factors in our proposed trust model.

\end{itemize}
\section{Conclusion and Future Work}\label{sec:conclusion}
This paper proposes a QoS-based trust evaluation model and architecture for black-box data services. In the model, trust evaluation is based on service performance and data quality. Service performance is evaluated considering a weighted addition of the response time, the task success ratio, and the service's availability. 

Data quality is calculated as a correlation of data timeliness and database timeliness. The first result based on a data quality observability protocol shows the advantage to take both data quality and service performance in the service trust calculation. 
This paper introduces DETECT, which implements the trust evaluation model. We used an e-health scenario implemented on top of our system DETECT to validate the feasibility of the proposed model.  


In future work, we intend to enrich DETECT with a data quality evaluation proposal that allows to guess data timeliness and database timeliness without having access to data quality features with full observability.
We also intend to study the effect of the variation of the trust factors over the ranking of services by performing multiple requests over time to show the effect of the variation of the performance level and the data quality level of services on their trust ranking. This situation can be tested by injecting faults in services at different moments to decrease their data quality or performance. Finally, as discussed in the experimental results section, we need first to study the correlation/dependence that exists between the performance of given data service and its data quality and to enhance our trust model in a way to reduce/eliminate/delete the bias of the performance on the data quality factor. 

\section*{Acknowledgment}

Thanks to yy\footnote{to be precised later} for funding this project xx\footnote{to be precised later}.


\end{document}